\documentclass[aip, preprint, groupedaddresses,superscriptaddress]{revtex4-1} 

\usepackage{dcolumn}%
\usepackage{bm}%
\usepackage{graphicx,amssymb,amsmath}

\usepackage{color}

\begin{document}

\title{Bowl breakout, escaping the positive region when searching for saddle points}
\author{Andreas Pedersen}
\affiliation{Faculty of Physical Sciences, University of Iceland VR-III, 107 Reykjav\'{\i}k, Iceland}
\affiliation{Integrated Systems Laboratory, ETH Zurich, 8092 Zurich, Switzerland}
\author{Mathieu Luiser}
\affiliation{Integrated Systems Laboratory, ETH Zurich, 8092 Zurich, Switzerland}

\begin{abstract}
We present a scheme improving the minimum-mode following method for finding first order saddle points by confining the displacements of atoms to the subset of those subject to the largest force.
By doing so it is ensured that the displacement remains of a local character within regions where all eigenvalues of the Hessian matrix are positive. However, as soon as a region is entered where an eigenvalue turns negative all atoms are released to maintain the ability of determining concerted moves.
Applying the proposed scheme reduces the required number of force calls for the determination of connected saddle points by a factor two or more compared to a free search.
Furthermore, a wider distribution of the relevant low barrier saddle points is obtained.
Finally, the dependency on the initial distortion and the applied maximal step size is reduced making minimum-mode guided searches both more robust and applicable.
\end{abstract}

\maketitle

\section{Introduction}
\label{sec:introduction}


The ability to determine saddle points (SPs) on the potential energy surface (PES) has proven itself a valuable computational tool in areas such as catalysis\cite{Greeley:2004ia}, mobility of defects\cite{Pizzagalli:2008gy} and contaminants\cite{pedersen09_4036}, the mechanical behavior of nano structures\cite{Zhu:2007gpa}, and acceleration of dynamics simulations\cite{voter02_321} to mention a few. 
When investigating the PES the most relevant points are local minima where all the eigenvalues, $\lambda$, of the Hessian matrix of second order derivatives are positive. By analyzing the corresponding configurations,  structural features can be obtained that characterize the system at moderate temperatures.  
Similarly, first order SPs, which have a single negative eigenvalue, hold important information about the dynamical behavior of the system. Such points represent the lowest energy configurations on the enclosing hyperplane---the rim of the basin---from within a local minimization will converge onto the contained minimum.  
The nudged elastic band\cite{henkelman00_9901} (NEB) method is an efficient and well-tested approach to determine such SPs when both the initial and final minima are known. It is referred to as a double ended search. 

When only the initial minimum is known (single ended search) a popular group of methods utilizes the eigenmodes of the Hessian matrix to guide a climb upwards on the PES towards a SP. This approach was proposed by Cerjan and Miller\cite{Cerjan:1981ua} and is  referred to as mode following (MF) methods. The SPs surrounding a minimum are then sampled by following the eigenmodes one by one. However, as the direction of the eigenmode needs to be updated in every iteration, such schemes are in general only applicable to systems where an analytic expression for the Hessian matrix can be formulated. 
In later works it has been proposed always to follow the minimum-mode that corresponds to the lowest eigenvalue. These methods are referred to as minimum-mode following (MMF)
or hybrid eigenvector-following
and enables simulations of larger systems\cite{Henkelman:1999vr, munro99_3969, malek00_7723, munro99_3969}. 
The crucial difference between the original MF formulation and MMF comes from the fact that MMF only needs the minimum-mode that can be efficiently estimated, whereas the full Hessian matrix is required in MF.
Furthermore, with the MMF methods the search is guaranteed to converge onto a SP in the negative regions of the PES\cite{Pedersen:2011cd}. 
However, within positive regions of the PES a MMF search is problematic as the true minimum-mode corresponds to a zero-mode accelerating either translation or rotation. 
Translational zero-modes can easily be handled whereas the removal of rotational zero-modes is a complicated task as the rotational modes
in the general case
are tightly coupled to the atomic vibrations and an infinitesimal rotational acceleration needs to be assumed\cite{Wales:2003vn}. 
Breathing-modes, which are long-ranged and soft, may also cause computational problems. They are deceptive as they lead to high energy SPs despite their soft curvature.
It must be restated that the soft and zero-modes are only problematic in positive ($0\leq\lambda_{min}$) regions of the PES because as soon as the search enters a negative ($0>\lambda_{min}$) region the lowest eigenmode of the Hessian matrix by definition turns negative and the corresponding SP can be determined. 

When applying MMF it still remains a challenge to efficiently escape  bowl shaped regions of the PES where all eigenvalues are positive.
In the dimer method\cite{Henkelman:1999vr} the SP search is started from a configuration where a local distortion is applied to displace the system away from the minimum. To quickly escape the positive region, aggressive steps are conducted until the negative region is reached. It has been demonstrated that the widest SP distribution results from initial distortions created by elements drawn from a Gaussian distribution as compared to elements drawn from an even distribution or by displacing the system along the eigenmodes in the minimum\cite{Pedersen:2011cd}. 
%
%
Another technique consists of iteratively applying a constant displacement vector until the system enters the negative region as done in ART\cite{munro99_3969}. 
Specific regions of a structure can be carefully sampled by making  these subjects for the initial distortion.
%
K{\"a}stner and Sherwood\cite{Kastner:2008ct} proposed to apply a constant weighting matrix and thereby keep the SP search focused onto given parts of a molecule. By this rescaling of the forces a stronger 'signal' is obtained from the part of the structure that is under inspection. The modified force acting on the system affects the search both in the positive and negative regions, but as the force vanishes at the SP the imposed transformation does not affect where the SP is located.
If the bond structure of the system is known, a given bond can be iteratively stretched until the negative region is entered. Such a scheme resembles the idea behind the bond-boost method that constructs a boost potential in MD simulations\cite{miron03_6210}. For many molecular and surface systems the stretching of a specific bond closely resembles the true chemical process and the part of the SP search in the negative region is only a matter of obtaining a tighter convergence.  However, problems might occur for more complex structures where the stretching of single bonds and SP searches might not provide a sufficient sampling of the PES.
%
When simulating structures, where the regions of relevance are limited and well separated---such as crystalline defects---the active volume scheme~\cite{Xu:2011dg} can be applied. By doing so, larger systems can be addressed as the conducted SP searches are confined to predefined subsets of atoms. Furthermore, the scheme also greatly reduces the computational requirements to characterize new minima, as structural changes tend to be local and SPs within other regions remain valid.      
%
Finally, recent methods have enabled recycling of already determined SPs to compose new SPs for a given minimum\cite{Xu:2008jk,ElMellouhi:2008eq}. However, as such constructs do not accurately account for long ranged effects, these SPs must be reconverged in fully free searches that adapt them to the current minimum.

In this paper we propose a 'bowl breakout' scheme, which enables a mode-following SP search to more efficiently escape the positive region on the PES. By confining the search to the subset of atoms experiencing the largest forces the search is kept local within the positive regions of the PES. As a demonstrations of its applicability the 'bowl breakout' technique is combined with the minimum-mode following method (BB-MMF) and applied to two test systems. The coming Section~\ref{sec:methods} contains a detailed description of both MMF and 'bowl breakout', it is followed by results in Section~\ref{sec:results}, and finally conclusions are drawn in Section~\ref{sec:conclusions}. 

\section{Computational methods}
\label{sec:methods}
\subsection{Minimum-mode following method}
\label{sec:mmfm}

Guiding a SP search utilizing the eigenmodes of the Hessian matrix was initially suggested by Cerjan and Miller\cite{Cerjan:1981ua}. Rather than relying on the true force acting on the system they proposed to apply an effective force obtained by inverting the force components parallel to a given eigenmode of the Hessian matrix, which might be expressed as:
\begin{align}
\label{equ:mmfm1}
\mathbf{F}^{\textit{eff}}=\mathbf{F}-2(\mathbf{F}\cdot\mathbf{v})\mathbf{v},
\end{align}
where $\mathbf{v}$ is the normalized direction of the selected eigenmode and $\mathbf{F}$ is the true force acting on the system. 
In the vicinity of a SP such a mapping transforms an unstable first order SP into a stable minimum, provided that the eigenmode corresponding to the lowest eigenvalue is supplied. Within the surrounding region where the eigenvalue remains negative ordinary minimization methods converge onto the contained SP if the effective force is supplied.

Applying only the minimum-mode was introduced by Henkelman and J{\'o}nsson who used the 'dimer method'\cite{Henkelman:1999vr},
Lindsey and Wales used a variational principle\cite{munro99_3969}, while Malek and Mousseau applied the Lanczos method in their 'ART' scheme\cite{malek00_7723} to estimate the minimum-mode. All techniques rely only on forces and 
these usually only need to be computed ten times
to give a sufficiently accurate estimation of the lowest eigenmode. This has to be compared with the $\propto N{^2}$ force evaluations required to construct the full Hessian matrix numerically, where $N$ is the number of atoms.

Two of the main settings to specify when applying MMF are the maximal allowed step size and how to reach negative regions: (i) the maximal step size, $\Delta_{max} $, should be an optimum between allowing large steps when minimizing forces while preventing overstepping to maintain the ability to converge onto the nearest local extrema; (ii) to reach negative regions different strategies can be applied as summarized in Section~\ref{sec:introduction}. 'Bowl breakout' relies on the introduction of a local distortion whose epicenter and radius, $R_{d}$, must be specified. This is similar to the original MMF method by Henkelman and J{\'o}nsson\cite{Henkelman:1999vr}. The standard deviation, $\sigma_{d}$, for the components of the distortions imposed on atoms is also required. 
A thorough determination of the optimal values for these  parameters imposes a significant computational effort that, however, does not assure a high yield of connected SPs.
To inspect how $R_{d}$ and $\Delta_{max}$ affect the performance of the SP search algorithm 
with and without the 'bowl breakout' scheme activated
three different sets have been selected:
\begin{enumerate}
  \item[a)] $R_{d}=2.5$~\AA\ and $\Delta_{max}=0.2$~\AA,
  \item[b)] $R_{d}=2.5$~\AA\ and $\Delta_{max}=0.4$~\AA,
  \item[c)] $R_{d}=7.5$~\AA\ and $\Delta_{max}=0.2$~\AA,
\end{enumerate}
where in a) a gentle initial distortion and a conservative step size are used; b) represents a gentle distortion, but an aggressive step size.  Finally c) introduces a large distortion, but a conservative step size. In all cases $\sigma_{d}$ is 0.1~\AA\ and the distortion is applied to the $x$, $y$ and $z$ coordinates of the atoms that are disturbed.

For calculations relying on DFT or other self-consistent procedures to obtain the electronic degrees of freedom (EDOF) and the atomic interactions b) is likely too aggressive for the MF methods to remain stable unless an unreasonably strict convergence criteria is imposed when calculating the EDOF. From a small test batch applying DFT the initial distortion labelled c) appears too large since all the searches start in high energy regions on the PES ($>20$~eV), which causes an immediate termination of the search process, see Section~\ref{sec:results} for definitions. 

\subsection{Bowl breakout}
\begin{figure}
\includegraphics[width=.8\textwidth]{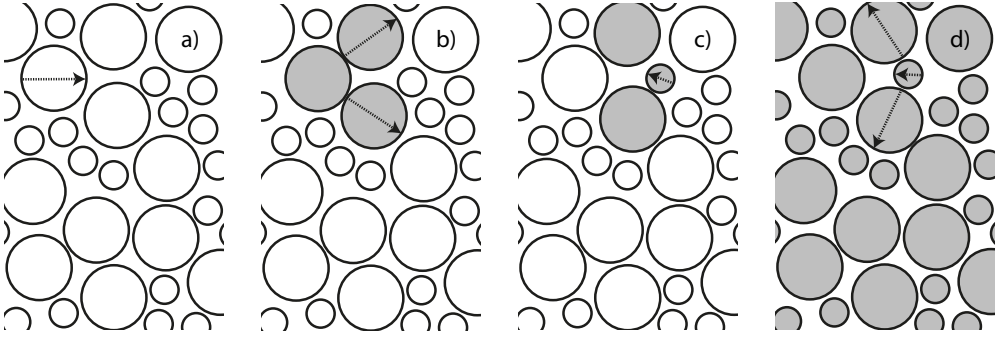}
\caption{
Schematic of the 'bowl breakout' method. a) An initial distortion is induced on the structure marked by the dashed arrow on the atom in the upper left corner. To escape the positive region and converge onto low laying saddle points only the atoms subject to the largest forces are displaced. Here a confinement of three atoms, $N_{\textit{confine}}=$ 3, is imposed until the negative region is reached. The grey shaded atoms are the ones undergoing displacements. Subplot d) represents the point where the negative region is reached and all atoms become free to move. It should be noted that the confinement adapt itself to a given configuration and is capable of following a changing set of active atoms as indicated in b) and c) while keeping the displacements local. 
}
\label{fig:confine_1}
\end{figure}
\begin{figure}
\includegraphics[width=.4\textwidth]{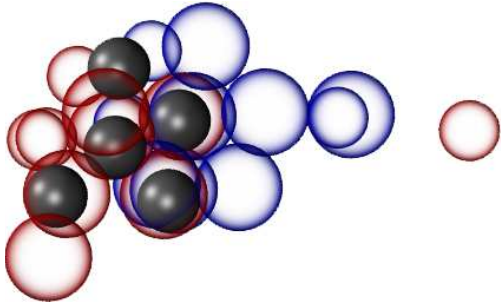}
\caption{
(Color online) Cutout from the amorphous CuAg alloy showing the set of atoms that are subject to the largest force when the saddle point search is initiated (blue) and when the search exits the negative region (red). A confinement of ten atoms is applied ($N_{\textit{confine}}=$ 10). The five atoms that have undergone the largest displacements to reach the saddle point configuration ($>0.6$~\AA) are shown as solid grey with a reduced radius. Large atoms are Ag and smaller ones are Cu. 
To obtain this configuration a barrier of 1.20~eV must be climbed, which is within the range of high barrier events  for the given minimum.
The Ag atoms are subjects to the largest displacements.  
}
\label{fig:confine_2}
\end{figure}
As an improvement for efficiently reaching the negative regions it is proposed to confine displacements to a subset of atoms within positive regions. 
To preserve the ability of the system to adapt itself to ongoing changes the members of this subset are updated in every iteration and thereby enclose the most relevant parts of the structure. The scheme for choosing the subset members consist in extracting the $N_{\textit{confine}}$ atoms subject to the largest forces, which ensures that the atoms being  mostly displaced from the local low energy configuration are included keeping the climb towards a SP local. Applying this scheme the atoms included in the confined search must fulfill:
\begin{align}
\label{equ:cc}
\text{Confinement} =
  	\begin{cases} 
  		True  & \text{ if \: $|\mathbf{F}_i| \geq |\mathbf{F}|_{confine}$} \\
   		Flase & \text{ if \: $|\mathbf{F}_i| < |\mathbf{F}|_{confine}$}
	\end{cases}
\end{align}
where $|\mathbf{F}_i|$ is the norm of the force acting on atom $i$, and $|\mathbf{F}|_{confine}$ represents the atom within the subset that has the smallest norm. As the latter value varies from configuration to configuration it needs to be determined for each structure visited. 
Once this is done a displacement is imposed to the included atoms that follows the direction of the effective force acting on the system scaled by the length of the maximal allowed step size, $\Delta_{max}$, as described in Section~\ref{sec:mmfm}. 
It is important to note that the proposed confinement is released when an inflection point on the PES is passed and $\lambda_{min}$ changes sign (from positive to negative)
but reapplied if the system falls back into a positive region. The force discontinuity imposed when the crossing takes place is a potential source of frustration, however, for the simulated systems it is of no significance.

A schematic representation of the 'bowl breakout' method is given in Fig.~\ref{fig:confine_1}. In this example the confinement is a subset of three atoms ($N_{\textit{confine}}=$ 3). 
By applying the proposed method the SP search is unlikely to end up in a long-ranged mode as a consequence of keeping the imposed structural changes local. 
Throughout the search process the members of the subset changes, which underlines the adaptation capabilities of the method that focuses the upward climb onto a distorted local environments. 
Since the subset automatically adapts, no knowledge about structural features is required {\it a priori} and the method is generally applicable. The introduced computational overhead is insignificant because the required additional operations rely only on forces that need to be determined in any case for the given configuration. 

Fig.~\ref{fig:confine_2} illustrates the functionality of the method for a real system. What should be noted is that the local displacements remain local and are adapted throughout the search. 
The wide overlap between the set of atoms that are subject to the largest forces in the final positive configuration and the set of atoms mostly displaced in the SP configuration is also relevant.

\section{Results}
\label{sec:results}
\subsection{Saddle point searches}

Searches for SPs on the PES are done by employing tools from the eon software package\footnote{http://theory.cm.utexas.edu/eon/ version 2213}, which is a freely available collection of methods to simulate rare event systems\cite{Chill:2014ct}. By combining 'bowl breakout' and the existing MMF code SP searches are conducted imposing a confinement of either: $N_{\textit{confine}}=$ 1, 5, 10, 20, 40 or that all atoms are free. The radius for the initial distortion, $R_{d}$, and the allowed step, $\Delta_{max}$, are varied as described in Section~\ref{sec:mmfm}. For the simulated systems batches of 1000 searches are computed. 

The outcome of the individual SP searches is classified into three groups: {\it Good}, the search converges onto a SP that connects to 
the minimum from where the search was initiated within a range of 0.01~eV and 0.1~\AA\; {\it Not connected}, the search converges onto a SP, but neither of the two minima matches the initial minimum; {\it Bad}, in this case convergence is not obtained, which mainly occurs because the iteration limit (1000) is reached, or a too high region on the PES is visited ($>20$~eV as compared to the initial minimum). The remaining searches are unclassified, but constitute less than 1~\% of the total.
A high ratio of {\it Not connected} searches indicates that the PES has SPs that are close in configuration space. If the {\it Bad} searches are terminated because too many iterations are required the search tends to roam in the positive region. If a region with too high energy is visited, the search termination is likely to be caused by a too large initial distortion ($R_{d}$) or a too large maximum step ($\Delta_{max}$).

All reported numbers are the grand totals of required force calls (FC), which account for the SP searches as well as the subsequent determination of the two minima connected to these SPs. 
Furthermore, performance numbers for the standard case as defined on optbench.org site can be found on the respective web page. 

\subsection{Simulated systems}
In all simulated systems the atomic interactions are modeled by applying the effective medium theory\cite{Jacobsen:1996vw} (EMT) implemented in ASAP\footnote{https://wiki.fysik.dtu.dk/asap/} that is capable of describing several metals and alloys.   
The simulated supercells contain approximately a thousand atoms that  are subject to periodic boundary conditions along $x$, $y$ and $z$.  These relatively large systems are used to validate the capabilities of the method on a high dimensional PES.
To anneal the structures each system is prepared by relaxing the size of the box and the atomic degrees of freedom using molecular dynamics and SP traversals\cite{Plasencia:2014hc}, which result in low energy structures.

An ordered FCC copper crystal containing a vacancy is the first of two simulated systems. It is a representative for point defects in structures with a highly ordered bulk phase and of relevance because similar systems are mainly addressed when DFT is applied in the evaluation of the atomic interactions\cite{Xu:2010bw, Benediktsson:2013hx}. Besides this crystalline sample a disordered structure is simulated, which is a copper silver alloy in an amorphous phase. This system is chosen to challenge the method with a more complex PES.

 


\subsubsection{Vacancy in copper} 
\begin{figure}
\includegraphics[width=.8\textwidth]{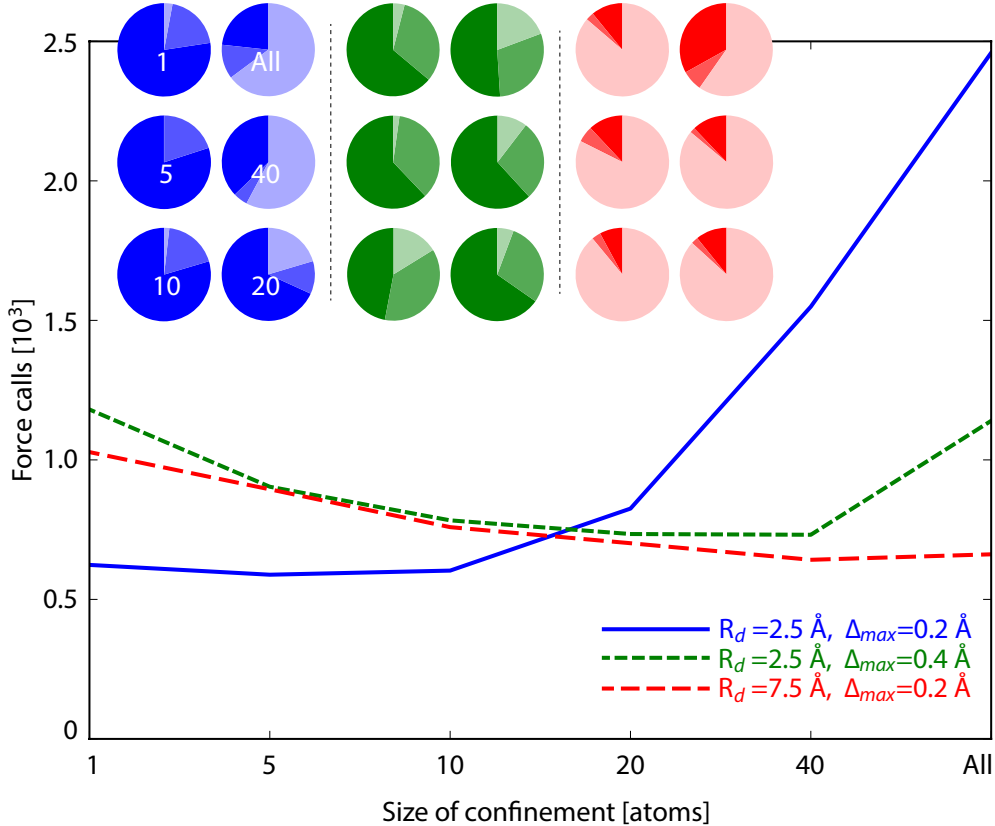}
\caption{
(Color online) Vacancy in a copper FCC lattice. The average number of force calls required per good saddle point is reported. Inset shows the distribution  of the located saddle points.  Blue corresponds to the parameter set a), green to set b) and red to set c) as defined in section~\ref{sec:methods}.  In the pie charts the darkest tones represent searches converging onto {\it Good} saddle points, the middle tones indicate {\it Not connected} saddle points, and the faintest tones refer to {\it Bad} searches. 
}
\label{fig:dist_vac_cu_3}
\end{figure}
\begin{figure}
\includegraphics[width=.8\textwidth]{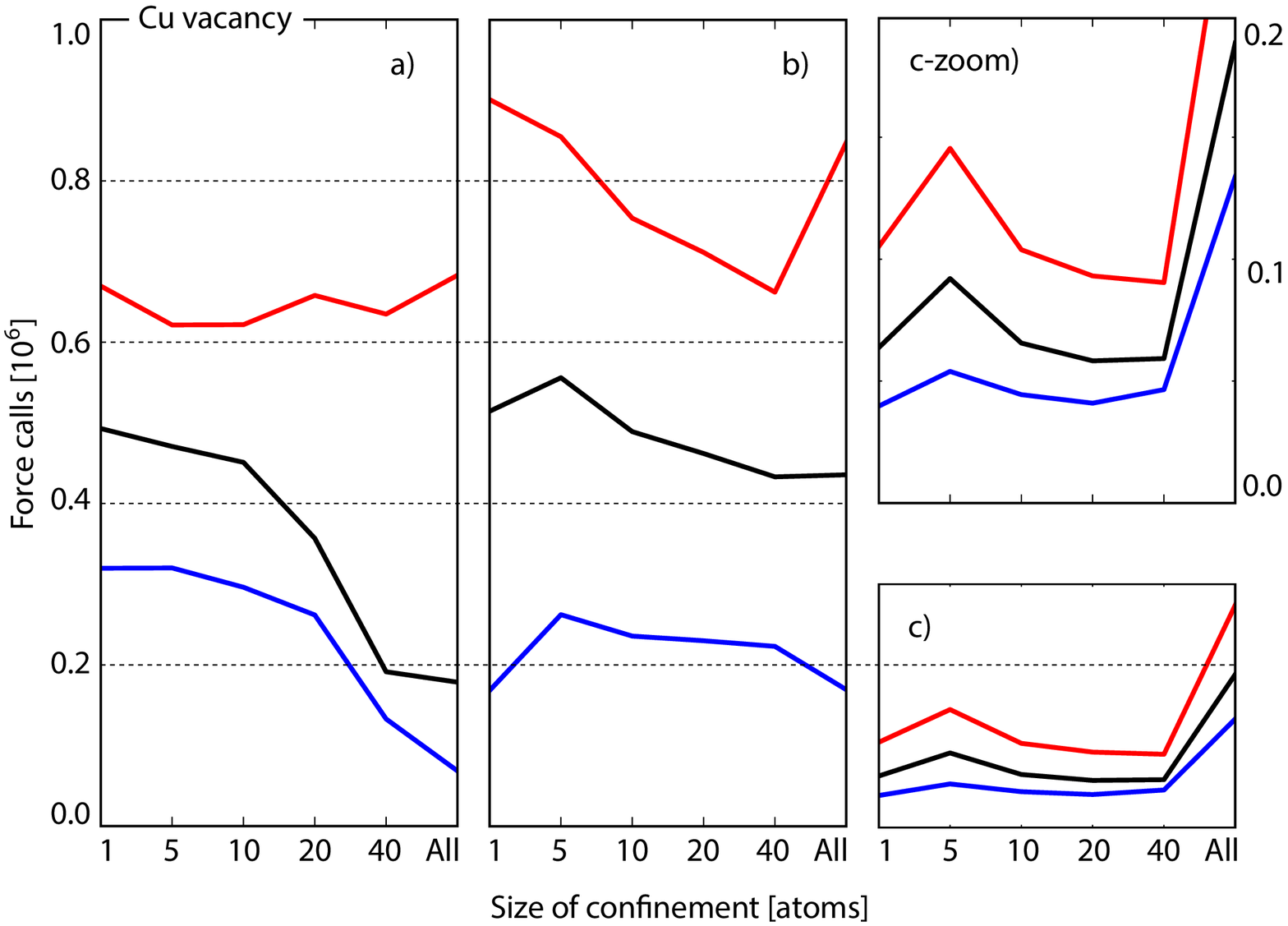}
\caption{
(Color online) Cumulative distribution of the force calls required by 1000 saddle point searches as a function of the applied confinement for a vacancy in bulk Cu. The force calls are sorted according to the outcome of the search.  Blue is for successful searches, black is for not connected and successful searches, and red is for bad searches, not connected and successful.
The leftmost column represents parameter set a), the center set b) and the rightmost set c) where the y-scale has been changed in the c-zoom.
}
\label{fig:fc_vac_cu_4}
\end{figure}
The vacancy in a Cu crystal is the simplest of the inspected system. 
In Fig.~\ref{fig:dist_vac_cu_3} the required number of FC for the different $N_{\textit{confine}}$ values and the ratio of {\it Good}, {\it Not connected} and {\it Bad} searches are reported.
The general observations are that approximately 1000 FC are required to determine and characterize a {\it Good} SP, in the following referred to as SP. It is important to note that this number accounts for all required FC including the searches that did not yield SPs. The best performance is obtained when applying parameter set a) and $N_{\textit{confine}}=$ 5 for which $\sim$600 FC are required per SP.  However, by applying scheme c) a similar performance is obtained for the fully free SP searches. 

When applying the conservative parameter set a) more than 75\% of the SP searches are successful when $N_{\textit{confine}}$ is equal or smaller than ten. The required number of FCs per SP in these cases is less than 600. When the size of the confinement increases and 20 or 40 atoms become free to move, a significant part of the searches get terminated because the iteration limit is reached. This is an unfortunate behavior as the searches waste a large number of FCs while roaming. Finally, when all atoms are released the ratio of SP becomes less than 25\% and 2400 FCs are required per SP. Furthermore, only 1/7 of the FCs go into searches yielding SPs, which appears in Fig.~\ref{fig:fc_vac_cu_4}. For the stricter confinements, where only ten or less atoms are free to move, half of the required FCs are lost in unsuccessful searches. 

The parameter set b) results in a significant decrease of SPs when the  confinement is imposed on ten atoms or less. However, the opposite behavior is seen when $N_{\textit{confine}}=$ 40 where the ratio of SPs increases by 25\% point. Irrespective of the size of the confinement, with parameter set b), the termination reason for {\it Bad} searches is not because the maximal number of iterations is reached but rather visits to high energy regions of the PES.

By applying the large initial distortion of parameter set c) the number of FCs per SP in the fully free searches decreases and reaches a level similar to the best results of the confined searches. However, the ratio of SPs for this parameter set is less than 30\%. For the confined searches the ration of SPs decreases even more dramatically so that less than 20\% of the confined searches result in SPs. Despite the high ratio of { \it Bad} searches the required number of FC does not significantly increase as most of these searches are immediately terminated because they are initiated in a region of the PES where the energy exceeds 20~eV.
 
\subsubsection{Amorphous copper silver alloy} 
\begin{figure}
\includegraphics[width=.8\textwidth]{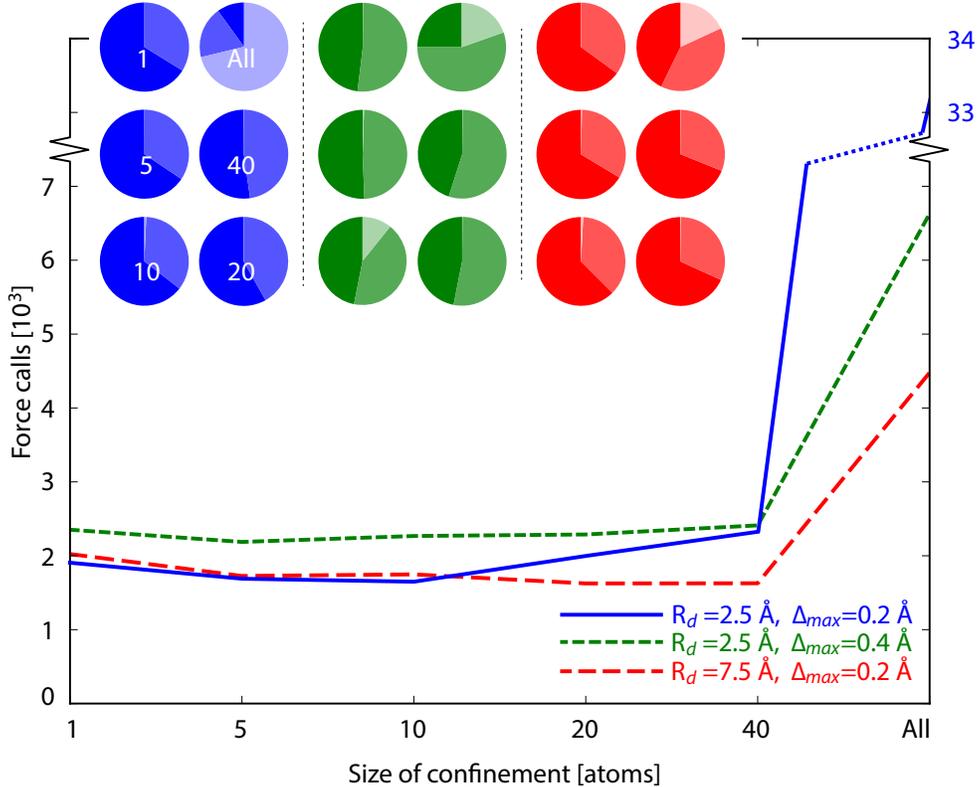}
\caption{
(Color online) Average number of force calls required per good saddle point and the distribution of the outcome of the saddle point searches for an amorphous CuAg alloy. The same color scheme as as in Fig.~\ref{fig:dist_vac_cu_3} is used.}
\label{fig:fc_cuag_5}
\end{figure}
\begin{figure}
\includegraphics[width=.8\textwidth]{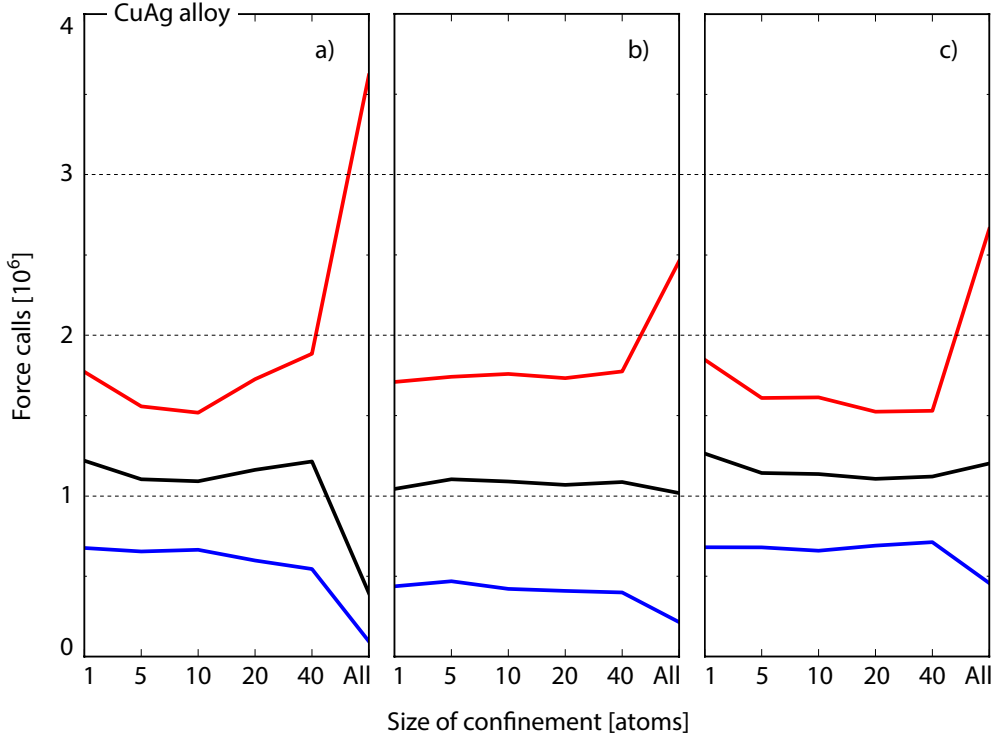}
\caption{
(Color online) Same as Fig.~\ref{fig:fc_vac_cu_4}, but for the amorphous CuAg alloy. 
}
\label{fig:fc_all_cuag_6}
\end{figure}
For the CuAg system an average of $\sim$2000~FCs is required to determine and characterize a SP when a confinement is imposed, see Fig.~\ref{fig:fc_cuag_5}. By comparing the results of the fully free searches to the results for the Cu crystal it appears that the PES for the CuAg alloy is more complicated for the MMF method to handle. As a consequence a significant performance gain is obtained by applying 'bowl breakout', which in all cases reduces the required FCs per SP by more than 50\%. The best performance is obtained by either conducting the SP search with $N_{\textit{confine}}=$ 10 and applying a gentle initial distortion, or by applying a large distortion but increasing the size of the imposed confinement to 40 atoms.

Using the parameter set a) $\sim$60\% of the confined searches result in SPs, but the remaining searches still converges. However, these SPs are not connected according to the requirements stated in Section~\ref{sec:results}. The results for the unconfined searches show that only one tenth of the searches result in SPs. The vast majority  (70\%) of the remaining searches are either terminated because a too high energy region on the PES is visited, or even worse, because the  iteration limit is reached. Fig.~\ref{fig:fc_all_cuag_6} indicates that the majority were indeed terminated for the latter reason as a large number of FC is required.
 
Allowing larger steps during the search significantly decreases the ratio of SP  when a confinement is imposed.  The higher ratio of unsuccessful searches for the parameter set b) has a direct effect on the average number of FCs required to determine a SP. An increase of $\sim$400 FCs as compared to a) is observed when $N_{\textit{confine}} \leq 10$. However, the negative impact depends on the size of the confined search and decreases as more atoms are included in the subset of atoms that are free to move. As for the crystalline structure the ratio of SPs increases and reaches a level of 25\% when all atoms are free to move.  As compared to parameter set a) a large part of the undesired population of {\it Bad} searches changes and becomes {\it Not connected} SPs, which reduces the computational demand to locate SPs by almost one order of magnitude. 
 
For the final batch c) where a large initial distortion is applied a high ratio of SPs is regained for the confined searches. Furthermore, the ratio of good searches for the largest values of $N_{\textit{confine}}$ and fully free searches exceeds the value of the two other parameter sets. As a result the smallest number of required FCs is obtained with a confinement to 20 atoms, as shown Fig.~\ref{fig:fc_cuag_5} where the red curve reaches the lowest level when $N_{\textit{confine}}=$ 20 and 40. 
When compared to the results for the copper vacancy sample the largest qualitative difference appear for parameter set c) where a large initial distortion is highly beneficial when all atoms are free. To understand this it must be noticed that the vacancy system is ordered with a homogenous and regular PES, which enables an efficient dissipation of unfavorable distortions through long ranged collective rearrangements. For such  restructuring to take place all atoms are required to be free to move and will be hindered by the confinement. However, for the more general case with a distribution of both bond lengths and binding energies---amorphous CuAg alloy---the resulting PES is rich in local features and the fully free search loses it advantage.


\subsection{Distribution of the determined saddle points} 
\begin{figure}
\includegraphics[width=.8\textwidth]{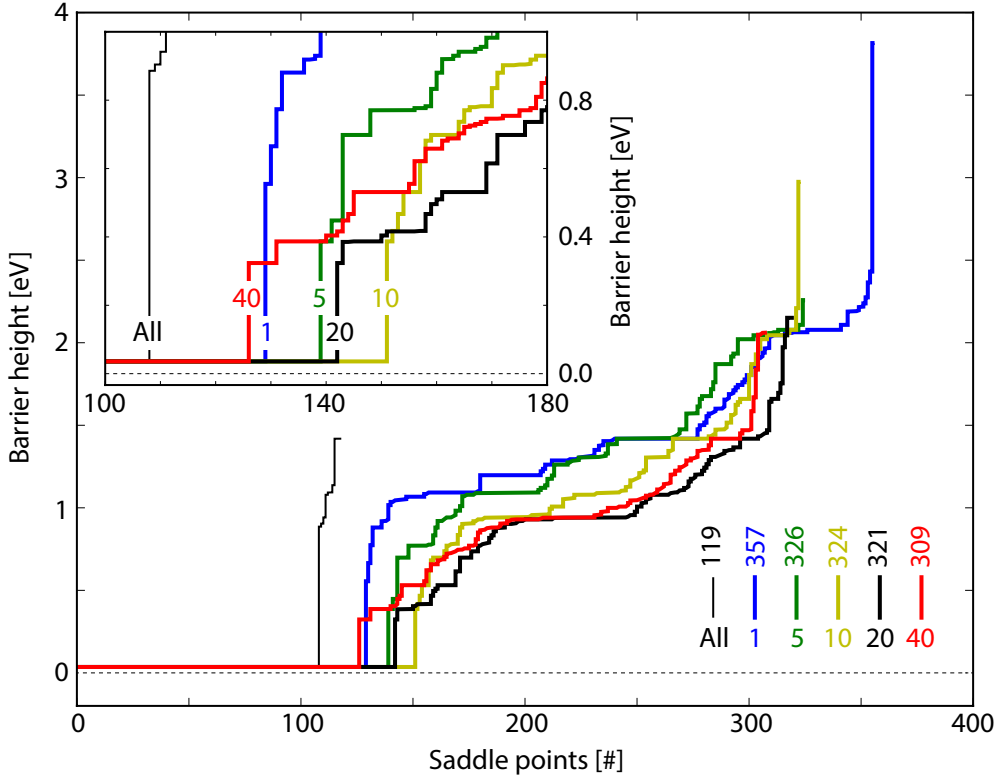}
\caption{
(Color online) Distribution of the barrier heights located within a batch of 500 saddle point  searches for the amorphous CuAg sample initiated from the region containing the lowest barrier event using parameter set a).  The numbers listed above the legends (119, 357, ... 309) are the number of good saddle points while the numbers below (All, 1, ... 40) represent the size of the confined search, $N_{\textit{confine}}$.  In all cases the lowest barrier event involving a barrier of 0.03~eV is determined by more than 20~\% of the searches. The inset shows that by applying confined searches several other low barrier events are determined. Furthermore, the confined searches also yield a significantly large variety of saddle points that appears as steps on the curves.  
}
\label{fig:cuag_barriers_7}
\end{figure}
To inspect how confined searches affect the distribution of SPs additional batches of 500 searches are conducted for the CuAg system using the parameter set a). The obtained distributions should be comparable as all searches are initiated by distorting a specific point in the structure, which is centered on the mostly displaced atom in the determined event with the lowest barrier. For this event to take place the system must overcome a barrier of 0.03~eV and the most active atom is displaced by more than 0.8~\AA\ before reaching the SP configuration.  
The distribution of the barriers for the determined events is shown in Fig.~\ref{fig:cuag_barriers_7}. In the figure it should be noted that confined searches outperform free searches both in terms of the number of times the lowest barrier event is determined, the total number of SPs, and the width of the distributions of the located SPs that also is apparent in Tab.~\ref{tab:sp_determination}. 

From the inset in Fig.~\ref{fig:cuag_barriers_7} one can see that the SP corresponding to the second lowest event, which involves a barrier of 0.32~eV, only gets determined when 
$N_{\textit{confine}} \geq 20$. A closer analysis of this event reveals that two atoms move by more than 2~\AA, four atoms move by more than 1~\AA\ (two when the atoms moving by more than 2~\AA\ are excluded), and sixteen more than 0.5~\AA\ (twelve when more active atoms are excluded). 
A possible explanation as to why the second lowest event is not determined by when with a smaller sub-set of atoms ($N_{\textit{confine}}< 20$) is applied might be that mainly atoms distant from the initial distortion are involved in the event. Since searches with a small confinement are kept local, they cannot reach eventual displacements at longer distances. Another less desirable possibility is that the considered event is a concerted move involving too many atoms for the stricter confinement to allow the system to make the required displacements. 
To determine the exact cause of the problem a final series of searches is conducted that is initiated from the mostly displaced atom for this event. From the lower section of Table~\ref{tab:sp_determination} it turns out that when $N_{\textit{confine}}$ is restricted to 5 atoms or less the second lowest SP is only determined in four out of the 500 searches or less, which indicates a limited ability of the 'bowl breakout' to determine concerted events when a too small confinement is applied. Nevertheless, when the confinement includes 10 or more atoms the concerted event is determined as frequently as for the free searches, but the number of unique SPs increases by a factor of six.      

\begin {table}
\begin{tabular}{ l || r | r | r | r | r | r }
$N_{\textit{confine}}$     & 1  & 5 & 10 & 20 & 40 & All\\
\hline  
Lowest	event & 129 & 139 & 151 & 142 & 126 & 108 \\
Unique events & 82  & 67  & 77  & 81  & 93  & 8 \\
\hline  
Second event  & 0  & 4  & 19 & 37 & 56 & 14  \\
Unique events & 85 & 86 & 91 & 83 & 93 & 15

\end{tabular} 
\caption{
Determination frequency of the two lowest barrier saddle points and the number of unique event within batches of 500 searches. All the searches are initiated from the atom that is mostly displaced in the saddle point  configuration for the given event as compared to the initial minimum.
}
\label{tab:sp_determination}
\end{table}
%

It should be mentioned that low barrier events such as the one (0.03~eV) mentioned above might be problematic if the determined SPs are supplied to an event table for adaptive kinetic Monte Carlo\cite{Henkelman:2001ur} simulations. However, when only a limited number of such low barrier events exist they can be efficiently handled by applying coarse graining\cite{Pedersen:2012wq}. In the present case the minimum, which is connected by the low barrier event, is observed to have a similar SP distribution as the initial minimum and the reverse mechanism back to this state is also a low barrier event. Since the two minima are separated by a low energy barrier and are close in configuration space the observed high ratio of {\it Not connected} searches in Fig.~\ref{fig:fc_cuag_5} is likely to be caused by SPs that belong to the neighbor minimum.

\section{Conclusions}
\label{sec:conclusions}
A method for escaping the positive region on the potential energy surface when conducting searches for saddle points is proposed. By confining the imposed displacements to the subset of atoms that are subject to the largest forces within positive regions it is possible to reduce the influence of global-modes. Confined searches are imposed as long as all eigenvalues of the Hessian matrix are positive. As soon as a negative region is reached all atoms regain their freedom and the system can converge onto saddle points representing concerted displacements of many atoms.

Applying the proposed method improves the performance of the mimimum-mode following method in the following ways: (i) a higher ratio of connected saddle points---as compared to fully free searches---is obtained. For a disordered structure this ratio is improved by more than 80\%; (ii) the distribution of unique saddle points resulting within a given number of searches widens by a factor six or more; (iii) finally, the occurrence of searches, which are terminated because too many iteration are conducted, is reduced. Combined, these improvements result in a reduction of 50\% in the number of required force calls per connected saddle point for the considered disordered structure.
Besides enhancing performance the applicability of the mimimum-mode following method is also improved as a relatively homogenous behavior is obtained when imposing a confinement in positive regions. Furthermore, the dependence on the parameters for the initial distortion and the maximal step size is weakened. To apply the proposed scheme the only additional parameter to supply is the size of the confinement. 

For the applied empirical potential the best performance is obtained by imposing a large initial distortion. However, when atomic interactions are modeled by DFT large distortions are observed to bring the system into regions of high energies and a more gentle approach seems to be necessary.    
Searches confined to 20 atoms appear to be the best compromise between performance in terms of force calls per good saddle point and the width of the distribution for the determined saddle points.
%



\acknowledgements
The authors would like to acknowledge ERC for funding obtained from the starting grant (E-MOBILE), the Icelandic Research Fund (RANNIS) and the University of Iceland Research Fund. They would also like to thank Hannes J{\'o}nsson at the University of Iceland for valuable inputs and fruitful discussions.

\bibliography{confine,confine_special}

\begin{thebibliography}{26}%
\makeatletter
\providecommand \@ifxundefined [1]{%
 \@ifx{#1\undefined}
}%
\providecommand \@ifnum [1]{%
 \ifnum #1\expandafter \@firstoftwo
 \else \expandafter \@secondoftwo
 \fi
}%
\providecommand \@ifx [1]{%
 \ifx #1\expandafter \@firstoftwo
 \else \expandafter \@secondoftwo
 \fi
}%
\providecommand \natexlab [1]{#1}%
\providecommand \enquote  [1]{``#1''}%
\providecommand \bibnamefont  [1]{#1}%
\providecommand \bibfnamefont [1]{#1}%
\providecommand \citenamefont [1]{#1}%
\providecommand \href@noop [0]{\@secondoftwo}%
\providecommand \href [0]{\begingroup \@sanitize@url \@href}%
\providecommand \@href[1]{\@@startlink{#1}\@@href}%
\providecommand \@@href[1]{\endgroup#1\@@endlink}%
\providecommand \@sanitize@url [0]{\catcode `\\12\catcode `\$12\catcode
  `\&12\catcode `\#12\catcode `\^12\catcode `\_12\catcode `\%12\relax}%
\providecommand \@@startlink[1]{}%
\providecommand \@@endlink[0]{}%
\providecommand \url  [0]{\begingroup\@sanitize@url \@url }%
\providecommand \@url [1]{\endgroup\@href {#1}{\urlprefix }}%
\providecommand \urlprefix  [0]{URL }%
\providecommand \Eprint [0]{\href }%
\providecommand \doibase [0]{http://dx.doi.org/}%
\providecommand \selectlanguage [0]{\@gobble}%
\providecommand \bibinfo  [0]{\@secondoftwo}%
\providecommand \bibfield  [0]{\@secondoftwo}%
\providecommand \translation [1]{[#1]}%
\providecommand \BibitemOpen [0]{}%
\providecommand \bibitemStop [0]{}%
\providecommand \bibitemNoStop [0]{.\EOS\space}%
\providecommand \EOS [0]{\spacefactor3000\relax}%
\providecommand \BibitemShut  [1]{\csname bibitem#1\endcsname}%
\let\auto@bib@innerbib\@empty
\bibitem [{\citenamefont {Greeley}\ and\ \citenamefont
  {Mavrikakis}(2004)}]{Greeley:2004ia}%
  \BibitemOpen
  \bibfield  {author} {\bibinfo {author} {\bibfnamefont {J.}~\bibnamefont
  {Greeley}}\ and\ \bibinfo {author} {\bibfnamefont {M.}~\bibnamefont
  {Mavrikakis}},\ }\bibfield  {title} {\enquote {\bibinfo {title} {{Alloy
  catalysts designed from first principles}},}\ }\href@noop {} {\bibfield
  {journal} {\bibinfo  {journal} {Nature Materials}\ }\textbf {\bibinfo
  {volume} {3}},\ \bibinfo {pages} {810--815} (\bibinfo {year}
  {2004})}\BibitemShut {NoStop}%
\bibitem [{\citenamefont {Pizzagalli}\ \emph {et~al.}(2008)\citenamefont
  {Pizzagalli}, \citenamefont {Pedersen}, \citenamefont {Arnaldsson},
  \citenamefont {J{\'o}nsson},\ and\ \citenamefont
  {Beauchamp}}]{Pizzagalli:2008gy}%
  \BibitemOpen
  \bibfield  {author} {\bibinfo {author} {\bibfnamefont {L.}~\bibnamefont
  {Pizzagalli}}, \bibinfo {author} {\bibfnamefont {A.}~\bibnamefont
  {Pedersen}}, \bibinfo {author} {\bibfnamefont {A.}~\bibnamefont
  {Arnaldsson}}, \bibinfo {author} {\bibfnamefont {H.}~\bibnamefont
  {J{\'o}nsson}}, \ and\ \bibinfo {author} {\bibfnamefont {P.}~\bibnamefont
  {Beauchamp}},\ }\bibfield  {title} {\enquote {\bibinfo {title} {{Theoretical
  study of kinks on screw dislocation in silicon}},}\ }\href@noop {} {\bibfield
   {journal} {\bibinfo  {journal} {Physical Review B}\ }\textbf {\bibinfo
  {volume} {77}} (\bibinfo {year} {2008})}\BibitemShut {NoStop}%
\bibitem [{\citenamefont {Pedersen}\ and\ \citenamefont
  {J{\'o}nsson}(2009)}]{pedersen09_4036}%
  \BibitemOpen
  \bibfield  {author} {\bibinfo {author} {\bibfnamefont {A.}~\bibnamefont
  {Pedersen}}\ and\ \bibinfo {author} {\bibfnamefont {H.}~\bibnamefont
  {J{\'o}nsson}},\ }\bibfield  {title} {\enquote {\bibinfo {title}
  {{Simulations of hydrogen diffusion at grain boundaries in aluminum}},}\
  }\href@noop {} {\bibfield  {journal} {\bibinfo  {journal} {Acta Materialia}\
  }\textbf {\bibinfo {volume} {57}},\ \bibinfo {pages} {4036--4045} (\bibinfo
  {year} {2009})}\BibitemShut {NoStop}%
\bibitem [{\citenamefont {Zhu}\ \emph {et~al.}(2007)\citenamefont {Zhu},
  \citenamefont {Li}, \citenamefont {Samanta}, \citenamefont {Kim},\ and\
  \citenamefont {Suresh}}]{Zhu:2007gpa}%
  \BibitemOpen
  \bibfield  {author} {\bibinfo {author} {\bibfnamefont {T.}~\bibnamefont
  {Zhu}}, \bibinfo {author} {\bibfnamefont {J.}~\bibnamefont {Li}}, \bibinfo
  {author} {\bibfnamefont {A.}~\bibnamefont {Samanta}}, \bibinfo {author}
  {\bibfnamefont {H.~G.}\ \bibnamefont {Kim}}, \ and\ \bibinfo {author}
  {\bibfnamefont {S.}~\bibnamefont {Suresh}},\ }\bibfield  {title} {\enquote
  {\bibinfo {title} {{Interfacial plasticity governs strain rate sensitivity
  and ductility in nanostructured metals}},}\ }\href@noop {} {\bibfield
  {journal} {\bibinfo  {journal} {Proceedings of the National Academy of
  Sciences}\ }\textbf {\bibinfo {volume} {104}},\ \bibinfo {pages} {3031--3036}
  (\bibinfo {year} {2007})}\BibitemShut {NoStop}%
\bibitem [{\citenamefont {Voter}, \citenamefont {Montalenti},\ and\
  \citenamefont {Germann}(2002)}]{voter02_321}%
  \BibitemOpen
  \bibfield  {author} {\bibinfo {author} {\bibfnamefont {A.~F.}\ \bibnamefont
  {Voter}}, \bibinfo {author} {\bibfnamefont {F.}~\bibnamefont {Montalenti}}, \
  and\ \bibinfo {author} {\bibfnamefont {T.~C.}\ \bibnamefont {Germann}},\
  }\bibfield  {title} {\enquote {\bibinfo {title} {{Extending the Time Scale in
  Atomistic Simulation of Materials}},}\ }\href@noop {} {\bibfield  {journal}
  {\bibinfo  {journal} {Annual Review of Materials Research}\ }\textbf
  {\bibinfo {volume} {32}},\ \bibinfo {pages} {321--346} (\bibinfo {year}
  {2002})}\BibitemShut {NoStop}%
\bibitem [{\citenamefont {Henkelman}, \citenamefont {Uberuaga},\ and\
  \citenamefont {J{\'o}nsson}(2000)}]{henkelman00_9901}%
  \BibitemOpen
  \bibfield  {author} {\bibinfo {author} {\bibfnamefont {G.}~\bibnamefont
  {Henkelman}}, \bibinfo {author} {\bibfnamefont {B.}~\bibnamefont {Uberuaga}},
  \ and\ \bibinfo {author} {\bibfnamefont {H.}~\bibnamefont {J{\'o}nsson}},\
  }\bibfield  {title} {\enquote {\bibinfo {title} {{A climbing image nudged
  elastic band method for finding saddle points and minimum energy paths}},}\
  }\href@noop {} {\bibfield  {journal} {\bibinfo  {journal} {Journal of
  Chemical Physics}\ }\textbf {\bibinfo {volume} {113}},\ \bibinfo {pages}
  {9901--9904} (\bibinfo {year} {2000})}\BibitemShut {NoStop}%
\bibitem [{\citenamefont {Cerjan}\ and\ \citenamefont
  {Miller}(1981)}]{Cerjan:1981ua}%
  \BibitemOpen
  \bibfield  {author} {\bibinfo {author} {\bibfnamefont {C.~J.}\ \bibnamefont
  {Cerjan}}\ and\ \bibinfo {author} {\bibfnamefont {W.~H.}\ \bibnamefont
  {Miller}},\ }\bibfield  {title} {\enquote {\bibinfo {title} {{On Finding
  Transition-States}},}\ }\href@noop {} {\bibfield  {journal} {\bibinfo
  {journal} {Journal of Chemical Physics}\ }\textbf {\bibinfo {volume} {75}},\
  \bibinfo {pages} {2800--2806} (\bibinfo {year} {1981})}\BibitemShut {NoStop}%
\bibitem [{\citenamefont {Henkelman}\ and\ \citenamefont
  {J{\'o}nsson}(1999)}]{Henkelman:1999vr}%
  \BibitemOpen
  \bibfield  {author} {\bibinfo {author} {\bibfnamefont {G.}~\bibnamefont
  {Henkelman}}\ and\ \bibinfo {author} {\bibfnamefont {H.}~\bibnamefont
  {J{\'o}nsson}},\ }\bibfield  {title} {\enquote {\bibinfo {title} {{A dimer
  method for finding saddle points on high dimensional potential surfaces using
  only first derivatives}},}\ }\href@noop {} {\bibfield  {journal} {\bibinfo
  {journal} {Journal of Chemical Physics}\ }\textbf {\bibinfo {volume} {111}},\
  \bibinfo {pages} {7010--7022} (\bibinfo {year} {1999})}\BibitemShut {NoStop}%
\bibitem [{\citenamefont {Munro}\ and\ \citenamefont
  {Wales}(1999)}]{munro99_3969}%
  \BibitemOpen
  \bibfield  {author} {\bibinfo {author} {\bibfnamefont {L.~J.}\ \bibnamefont
  {Munro}}\ and\ \bibinfo {author} {\bibfnamefont {D.~J.}\ \bibnamefont
  {Wales}},\ }\bibfield  {title} {\enquote {\bibinfo {title} {{Defect migration
  in crystalline silicon}},}\ }\href@noop {} {\bibfield  {journal} {\bibinfo
  {journal} {Physical Review B}\ }\textbf {\bibinfo {volume} {59}},\ \bibinfo
  {pages} {3969--3980} (\bibinfo {year} {1999})}\BibitemShut {NoStop}%
\bibitem [{\citenamefont {Malek}\ and\ \citenamefont
  {Mousseau}(2000)}]{malek00_7723}%
  \BibitemOpen
  \bibfield  {author} {\bibinfo {author} {\bibfnamefont {R.}~\bibnamefont
  {Malek}}\ and\ \bibinfo {author} {\bibfnamefont {N.}~\bibnamefont
  {Mousseau}},\ }\bibfield  {title} {\enquote {\bibinfo {title} {{Dynamics of
  lennard-jones clusters A characterization of the activation-relaxation
  technique}},}\ }\href@noop {} {\bibfield  {journal} {\bibinfo  {journal}
  {Physical Review E}\ }\textbf {\bibinfo {volume} {62}},\ \bibinfo {pages}
  {7723--7728} (\bibinfo {year} {2000})}\BibitemShut {NoStop}%
\bibitem [{\citenamefont {Pedersen}, \citenamefont {Hafstein},\ and\
  \citenamefont {J{\'o}nsson}(2011)}]{Pedersen:2011cd}%
  \BibitemOpen
  \bibfield  {author} {\bibinfo {author} {\bibfnamefont {A.}~\bibnamefont
  {Pedersen}}, \bibinfo {author} {\bibfnamefont {S.~F.}\ \bibnamefont
  {Hafstein}}, \ and\ \bibinfo {author} {\bibfnamefont {H.}~\bibnamefont
  {J{\'o}nsson}},\ }\bibfield  {title} {\enquote {\bibinfo {title} {{Efficient
  Sampling of Saddle Points with the Minimum-Mode Following Method}},}\
  }\href@noop {} {\bibfield  {journal} {\bibinfo  {journal} {SIAM Journal on
  Scientific Computing}\ }\textbf {\bibinfo {volume} {33}},\ \bibinfo {pages}
  {633} (\bibinfo {year} {2011})}\BibitemShut {NoStop}%
\bibitem [{\citenamefont {Wales}(2003)}]{Wales:2003vn}%
  \BibitemOpen
  \bibfield  {author} {\bibinfo {author} {\bibfnamefont {D.}~\bibnamefont
  {Wales}},\ }\href@noop {} {\emph {\bibinfo {title} {{Energy Landscapes}}}},\
  Applications to Clusters, Biomolecules and Glasses\ (\bibinfo  {publisher}
  {Cambridge University Press},\ \bibinfo {year} {2003})\BibitemShut {NoStop}%
\bibitem [{\citenamefont {K{\"a}stner}\ and\ \citenamefont
  {Sherwood}(2008)}]{Kastner:2008ct}%
  \BibitemOpen
  \bibfield  {author} {\bibinfo {author} {\bibfnamefont {J.}~\bibnamefont
  {K{\"a}stner}}\ and\ \bibinfo {author} {\bibfnamefont {P.}~\bibnamefont
  {Sherwood}},\ }\bibfield  {title} {\enquote {\bibinfo {title} {{Superlinearly
  converging dimer method for transition state search}},}\ }\href@noop {}
  {\bibfield  {journal} {\bibinfo  {journal} {The Journal of Chemical Physics}\
  }\textbf {\bibinfo {volume} {128}},\ \bibinfo {pages} {014106} (\bibinfo
  {year} {2008})}\BibitemShut {NoStop}%
\bibitem [{\citenamefont {Miron}\ and\ \citenamefont
  {Fichthorn}(2003)}]{miron03_6210}%
  \BibitemOpen
  \bibfield  {author} {\bibinfo {author} {\bibfnamefont {R.~A.}\ \bibnamefont
  {Miron}}\ and\ \bibinfo {author} {\bibfnamefont {K.~A.}\ \bibnamefont
  {Fichthorn}},\ }\bibfield  {title} {\enquote {\bibinfo {title} {{Accelerated
  molecular dynamics with the bond-boost method}},}\ }\href@noop {} {\bibfield
  {journal} {\bibinfo  {journal} {Journal of Chemical Physics}\ }\textbf
  {\bibinfo {volume} {119}},\ \bibinfo {pages} {6210--6216} (\bibinfo {year}
  {2003})}\BibitemShut {NoStop}%
\bibitem [{\citenamefont {Xu}, \citenamefont {Osetsky},\ and\ \citenamefont
  {Stoller}(2011)}]{Xu:2011dg}%
  \BibitemOpen
  \bibfield  {author} {\bibinfo {author} {\bibfnamefont {H.}~\bibnamefont
  {Xu}}, \bibinfo {author} {\bibfnamefont {Y.~N.}\ \bibnamefont {Osetsky}}, \
  and\ \bibinfo {author} {\bibfnamefont {R.~E.}\ \bibnamefont {Stoller}},\
  }\bibfield  {title} {\enquote {\bibinfo {title} {{Simulating complex
  atomistic processes: On-the-fly kinetic Monte Carlo scheme with selective
  active volumes}},}\ }\href@noop {} {\bibfield  {journal} {\bibinfo  {journal}
  {Physical Review B}\ }\textbf {\bibinfo {volume} {84}},\ \bibinfo {pages}
  {132103} (\bibinfo {year} {2011})}\BibitemShut {NoStop}%
\bibitem [{\citenamefont {Xu}\ and\ \citenamefont
  {Henkelman}(2008)}]{Xu:2008jk}%
  \BibitemOpen
  \bibfield  {author} {\bibinfo {author} {\bibfnamefont {L.}~\bibnamefont
  {Xu}}\ and\ \bibinfo {author} {\bibfnamefont {G.}~\bibnamefont {Henkelman}},\
  }\bibfield  {title} {\enquote {\bibinfo {title} {{Adaptive kinetic Monte
  Carlo for first-principles accelerated dynamics}},}\ }\href@noop {}
  {\bibfield  {journal} {\bibinfo  {journal} {The Journal of Chemical Physics}\
  }\textbf {\bibinfo {volume} {129}},\ \bibinfo {pages} {114104} (\bibinfo
  {year} {2008})}\BibitemShut {NoStop}%
\bibitem [{\citenamefont {El-Mellouhi}, \citenamefont {Mousseau},\ and\
  \citenamefont {Lewis}(2008)}]{ElMellouhi:2008eq}%
  \BibitemOpen
  \bibfield  {author} {\bibinfo {author} {\bibfnamefont {F.}~\bibnamefont
  {El-Mellouhi}}, \bibinfo {author} {\bibfnamefont {N.}~\bibnamefont
  {Mousseau}}, \ and\ \bibinfo {author} {\bibfnamefont {L.}~\bibnamefont
  {Lewis}},\ }\bibfield  {title} {\enquote {\bibinfo {title} {{Kinetic
  activation-relaxation technique: An off-lattice self-learning kinetic Monte
  Carlo algorithm}},}\ }\href@noop {} {\bibfield  {journal} {\bibinfo
  {journal} {Physical Review B}\ }\textbf {\bibinfo {volume} {78}},\ \bibinfo
  {pages} {153202} (\bibinfo {year} {2008})}\BibitemShut {NoStop}%
\bibitem [{Note1()}]{Note1}%
  \BibitemOpen
  \bibinfo {note} {Http://theory.cm.utexas.edu/eon/ version 2213}\BibitemShut
  {NoStop}%
\bibitem [{\citenamefont {Chill}\ \emph {et~al.}(2014)\citenamefont {Chill},
  \citenamefont {Welborn}, \citenamefont {Terrell}, \citenamefont {Zhang},
  \citenamefont {Berthet}, \citenamefont {Pedersen}, \citenamefont
  {J{\'o}nsson},\ and\ \citenamefont {Henkelman}}]{Chill:2014ct}%
  \BibitemOpen
  \bibfield  {author} {\bibinfo {author} {\bibfnamefont {S.~T.}\ \bibnamefont
  {Chill}}, \bibinfo {author} {\bibfnamefont {M.}~\bibnamefont {Welborn}},
  \bibinfo {author} {\bibfnamefont {R.}~\bibnamefont {Terrell}}, \bibinfo
  {author} {\bibfnamefont {L.}~\bibnamefont {Zhang}}, \bibinfo {author}
  {\bibfnamefont {J.-C.}\ \bibnamefont {Berthet}}, \bibinfo {author}
  {\bibfnamefont {A.}~\bibnamefont {Pedersen}}, \bibinfo {author}
  {\bibfnamefont {H.}~\bibnamefont {J{\'o}nsson}}, \ and\ \bibinfo {author}
  {\bibfnamefont {G.}~\bibnamefont {Henkelman}},\ }\bibfield  {title} {\enquote
  {\bibinfo {title} {{EON: software for long time simulations of atomic scale
  systems}},}\ }\href@noop {} {\bibfield  {journal} {\bibinfo  {journal}
  {Modelling and Simulation in Materials Science and Engineering}\ }\textbf
  {\bibinfo {volume} {22}},\ \bibinfo {pages} {055002} (\bibinfo {year}
  {2014})}\BibitemShut {NoStop}%
\bibitem [{\citenamefont {Jacobsen}, \citenamefont {Stoltze},\ and\
  \citenamefont {N{\o}rskov}(1996)}]{Jacobsen:1996vw}%
  \BibitemOpen
  \bibfield  {author} {\bibinfo {author} {\bibfnamefont {K.~W.}\ \bibnamefont
  {Jacobsen}}, \bibinfo {author} {\bibfnamefont {P.}~\bibnamefont {Stoltze}}, \
  and\ \bibinfo {author} {\bibfnamefont {J.~K.}\ \bibnamefont {N{\o}rskov}},\
  }\bibfield  {title} {\enquote {\bibinfo {title} {{A semi-empirical effective
  medium theory for metals and alloys}},}\ }\href@noop {} {\bibfield  {journal}
  {\bibinfo  {journal} {Surface Science}\ }\textbf {\bibinfo {volume} {366}},\
  \bibinfo {pages} {394--402} (\bibinfo {year} {1996})}\BibitemShut {NoStop}%
\bibitem [{Note2()}]{Note2}%
  \BibitemOpen
  \bibinfo {note} {Https://wiki.fysik.dtu.dk/asap/}\BibitemShut {NoStop}%
\bibitem [{\citenamefont {Plasencia}\ \emph {et~al.}(2014)\citenamefont
  {Plasencia}, \citenamefont {Pedersen}, \citenamefont {Berthet}, \citenamefont
  {J{\'o}nsson},\ and\ \citenamefont {Arnaldsson}}]{Plasencia:2014hc}%
  \BibitemOpen
  \bibfield  {author} {\bibinfo {author} {\bibfnamefont {M.}~\bibnamefont
  {Plasencia}}, \bibinfo {author} {\bibfnamefont {A.}~\bibnamefont {Pedersen}},
  \bibinfo {author} {\bibfnamefont {J.-C.}\ \bibnamefont {Berthet}}, \bibinfo
  {author} {\bibfnamefont {H.}~\bibnamefont {J{\'o}nsson}}, \ and\ \bibinfo
  {author} {\bibfnamefont {A.}~\bibnamefont {Arnaldsson}},\ }\bibfield  {title}
  {\enquote {\bibinfo {title} {{Computers {\&} Geosciences}},}\ }\href@noop {}
  {\bibfield  {journal} {\bibinfo  {journal} {Computers and Geosciences}\
  }\textbf {\bibinfo {volume} {65}},\ \bibinfo {pages} {110--117} (\bibinfo
  {year} {2014})}\BibitemShut {NoStop}%
\bibitem [{\citenamefont {Xu}\ and\ \citenamefont
  {Henkelman}(2010)}]{Xu:2010bw}%
  \BibitemOpen
  \bibfield  {author} {\bibinfo {author} {\bibfnamefont {L.}~\bibnamefont
  {Xu}}\ and\ \bibinfo {author} {\bibfnamefont {G.}~\bibnamefont {Henkelman}},\
  }\bibfield  {title} {\enquote {\bibinfo {title} {{Calculations of Li
  adsorption and diffusion on MgO(100) in comparison to Ca}},}\ }\href@noop {}
  {\bibfield  {journal} {\bibinfo  {journal} {Physical Review B}\ }\textbf
  {\bibinfo {volume} {82}},\ \bibinfo {pages} {115407} (\bibinfo {year}
  {2010})}\BibitemShut {NoStop}%
\bibitem [{\citenamefont {Benediktsson}\ \emph {et~al.}(2013)\citenamefont
  {Benediktsson}, \citenamefont {M{\'{y}}rdal}, \citenamefont {Maurya},\ and\
  \citenamefont {Pedersen}}]{Benediktsson:2013hx}%
  \BibitemOpen
  \bibfield  {author} {\bibinfo {author} {\bibfnamefont {M.~T.}\ \bibnamefont
  {Benediktsson}}, \bibinfo {author} {\bibfnamefont {K.~K.~G.}\ \bibnamefont
  {M{\'{y}}rdal}}, \bibinfo {author} {\bibfnamefont {P.}~\bibnamefont
  {Maurya}}, \ and\ \bibinfo {author} {\bibfnamefont {A.}~\bibnamefont
  {Pedersen}},\ }\bibfield  {title} {\enquote {\bibinfo {title} {{Stability and
  mobility of vacancy{\textendash}H complexes in Al}},}\ }\href@noop {}
  {\bibfield  {journal} {\bibinfo  {journal} {Journal of Physics-Condensed
  Matter}\ }\textbf {\bibinfo {volume} {25}},\ \bibinfo {pages} {375401}
  (\bibinfo {year} {2013})}\BibitemShut {NoStop}%
\bibitem [{\citenamefont {Henkelman}\ and\ \citenamefont
  {J{\'o}nsson}(2001)}]{Henkelman:2001ur}%
  \BibitemOpen
  \bibfield  {author} {\bibinfo {author} {\bibfnamefont {G.}~\bibnamefont
  {Henkelman}}\ and\ \bibinfo {author} {\bibfnamefont {H.}~\bibnamefont
  {J{\'o}nsson}},\ }\bibfield  {title} {\enquote {\bibinfo {title} {{Long time
  scale kinetic Monte Carlo simulations without lattice approximation and
  predefined event table}},}\ }\href@noop {} {\bibfield  {journal} {\bibinfo
  {journal} {Journal of Chemical Physics}\ }\textbf {\bibinfo {volume} {115}},\
  \bibinfo {pages} {9657--9666} (\bibinfo {year} {2001})}\BibitemShut {NoStop}%
\bibitem [{\citenamefont {Pedersen}, \citenamefont {Berthet},\ and\
  \citenamefont {J{\'o}nsson}(2012)}]{Pedersen:2012wq}%
  \BibitemOpen
  \bibfield  {author} {\bibinfo {author} {\bibfnamefont {A.}~\bibnamefont
  {Pedersen}}, \bibinfo {author} {\bibfnamefont {J.~C.}\ \bibnamefont
  {Berthet}}, \ and\ \bibinfo {author} {\bibfnamefont {H.}~\bibnamefont
  {J{\'o}nsson}},\ }\bibfield  {title} {\enquote {\bibinfo {title} {{Simulated
  annealing with coarse graining and distributed computing}},}\ }\href@noop {}
  {\bibfield  {journal} {\bibinfo  {journal} {Applied Parallel and Scientific
  Computing}\ ,\ \bibinfo {pages} {34--44}} (\bibinfo {year}
  {2012})}\BibitemShut {NoStop}%
\end{thebibliography}%
\end{document}